\documentclass[10pt, letterpaper, a4paper]{amsart}
\usepackage{amsmath}
\usepackage{latexsym}
\usepackage[all]{xy}
\usepackage{color}
\newtheorem{teorema}{Theorem}[section]
\newtheorem{definicion}[teorema]{Definition}
\include{amslatex}

\usepackage[margin=1.4in]{geometry}
\newtheorem{comentario}[teorema]{Remark}

\numberwithin{equation}{section}
\begin{document}
\begin{title}[On the notion of kinetic energy...]
{On the notion of kinetic energy in a spacetime of maximal proper acceleration}
\end{title}
\date{\today}
\maketitle
\begin{center}
\author{Ricardo Gallego Torrom\'e\footnote{Email: rigato39@gmail.com}}
\end{center}

\begin{abstract}
It is shown in this work that in spacetimes of maximal proper acceleration there are situations where a particle decelerates and simultaneously increases its kinetic energy. To illustrate the effect, a classical $1$-dimensional electric cavity-like scheme is discussed. The gain in kinetic energy by deceleration suggests an equivalence between velocity and acceleration from the energetic point of view. Such an equivalence is stated in a form resembling Born reciprocity principle but formulated in the framework of higher order jet spacetime geometry. We also show the existence in the theory of a dynamical superluminal sector for point particles with positive mass and characterized by superluminal speed and a proper acceleration larger than the maximal proper acceleration. Therefore, the speed of light and the maximal proper acceleration become only {\it separation conditions} at the classical level of the two dynamical sectors.
\end{abstract}

\section{Introduction}
First introduced by E. Caianiello in the context of his theory of {\it quantum geometry} \cite{Caianiello, Caianiello 1992}, the conjecture on the existence of a maximal acceleration has brought attention due to its deepness, ubiquity in theoretical physics and broad spectrum of phenomenological consequences \cite{Ricardo Nicolini 2018}. This fecundity is, however, overshadowed by its usual extremely large, attached value. Indeed, such large acceleration scales have precluded the possibility of direct observation of the phenomenological consequences of maximal proper acceleration, leaving the notion and its consequences almost in the realm of speculative theory. This situation is starting to change.  From one side, there are laser-plasma acceleration facilities and standard accelerator facilities capable of reaching acceleration scales where observable effects of a maximal proper acceleration in electrodynamics are accessible. From another side, there has been the theoretical development of certain models of electrodynamics suggesting a dramatic reduction of the value of the maximal proper acceleration for particle beams in accelerators \cite{Ricardo 2019, Ricardo 2024, Ricardo 2025, Ricardo 2025b}, making the conjecture to potential experimental falsifiability or verification.

The framework where such phenomenological consequences of maximal proper acceleration have been discussed is the {\it  theory of spacetimes of maximal proper acceleration}. The theory has its origin on the detailed analysis of three interrelated insights and problems. First, there is the observation that {\it the clock hypothesis} in relativistic theories, namely, the assumption that the proper time functional does not depend upon the proper acceleration of the path or on higher derivatives, fails for very large accelerations, as it happens in processes where radiation reaction effects are of relevance \cite{Mashhoon1990}. The clock hypothesis, which was assumed implicitly by Einstein in the development of theory of relativity and formulated explicitly in several occasions by Einstein himself and by Pauli \cite{Einstein1922, Pauli 1958}, is not theoretically justified except by its role in simplifying the mathematical constructions of the theory of relativity, leaving the feeling of being a principle not fundamentally necessary for the development of the theory. Second, the problem of the covariance of Caianiello's theory was a second motivation for our theory. A proposed solution to the problem was developed in \cite{Ricardo2007} by means of a theory of non-linear connections, but also interestingly and of ampler generality, that paper introduced very succinctly the relation between higher order jet geometry and classical measurement processes. The third issue related with the emergence of the theory of spacetimes of maximal proper acceleration was the issue of finding a consistent second order ordinary differential equation for a point charged particle without the pathologies of the Lorentz-Dirac equation. The resolution of this problem leaded to the notion of spacetime of maximal proper acceleration as a higher order jet geometry and to the notion of higher order jet fields \cite{Ricardo2012, Ricardo2015, Ricardo 2024, Ricardo 2025}. The physical significance of such constructions is that physical fields, spacetime geometry and matter loss their physical independence from each other and instead, the whole process of measurement and the ontological entities involved are unified in terms of higher order jet fields \cite{Ricardo2012, Saunders 1989, KolarMichorSlovak}. Moreover, besides the applications on electrodynamics and gravitation, the theory of spacetimes of maximal proper acceleration in its more general form, the theory of higher order jet geometry, finds a second natural application in Hamilton-Randers theory, an approach to emergent quantum mechanics where the fundamental degrees of freedom are constrained by a limit in proper acceleration and speed \cite{Ricardo05b,Ricardo06,Ricardo2014,Ricardo 2024b}.

In a spacetime of maximal proper acceleration, the kinetic energy and the momentum of a point particle not only depend upon the velocity of the particle, but also on its proper acceleration \cite{Ricardo 2024}. Consequently, the theory implies a modification of the inertial mass of a particle being accelerated in such a way that for an accelerated particle  there is a reinforcement of the inertial mass with respect to the mass measured by an inertial observer instantaneously at rest with the particle \cite{Ricardo 2024}. After discussing the notion of kinetic energy, we show in this article that the theory of spacetimes of maximal proper acceleration admits a {\it tachyonic sector} characterized by particles having positive mass but being super-luminal and having a proper acceleration beyond the proper acceleration limit. Despite the conditions for the physical realization of such dynamical sector are extreme, the novelty of the theory resides in that it applies to ordinary matter.
Second, it is shown that the more involved dependence of the kinetic energy of the particle in spacetimes of maximal proper acceleration the origin of the possibility of increasing the kinetic energy simultaneously as the particle decreases its speed, since the decrease in the kinetic energy due to a decrease in the velocity can be compensated or even surpassed by an increase in the energy due to an increase in the proper acceleration. Such an effect is illustrated by a toy model describing a point charged particle moving in a $1$-dimensional, classical, electric cavity. It is shown that when disregarding radiative effects and Bremsstrahlung, there is a non-trivial gain in energy by the particle when approaching a charged wire of the same charge sign. In addition, there are three Appendix. In the first one, we recall the notion of generalized higher order electrodynamics. In the second, we introduce an action functional for a point charged particle interacting with a higher order jet field in a spacetime of maximal proper acceleration and derive the equation of motion from this starting point. Note that additional phenomenological conditions are imposed during the derivation. In Appendix \ref{Appendix C}, the action is used to derive the general form of the stress-energy momentum tensor and a form of energy balance equation.

\section{Brief introduction to the notion of spacetime of maximal proper acceleration}
The theory of spacetimes of maximal proper acceleration assumes a classical scenario where smooth world lines are well defined. The proper time functional and the associated metric structure $g$ depend on the second jet of the test particle used to probe the spacetime.
Let $\vartheta:I\to M_4$ be a smooth curve describing the world line of a probe point particle and let $\,^2\vartheta(t):=\,\left(t,\vartheta^\mu(t),\frac{d{\vartheta}^\mu(t)}{dt},\frac{d^2{\vartheta}^\mu(t)}{dt^2}\right)$ be its second jet, an element of the second jet manifold $J^2M_4$ corresponding to the second jet lift of $\vartheta:I\to M_4$ at $\vartheta(t)$.

The metric $\eta$, of signature $(-1,1,1,1)$, is the metric structure in situations when the probe particle does not suffer acceleration. That is, at the point $x\in \,M_4$, the metric $\eta$ determines a bilinear form determined by the $1$-jet $\,^1\vartheta(t):=\,\left(t,\vartheta^\mu(t),\frac{d{\vartheta}^\mu(t)}{dt},\frac{d^2{\vartheta}^\mu(t)}{dt^2}=0\right)$ at $\vartheta(t)=\,x\in\,M_4$. Strictly speaking, this construction leads to a Finsler-Lorentzian type structure, but it is known how to obtain, by the averaging method  \cite{Ricardo 2021}, a Lorenztian structure from the aggregate of bilinear forms $\{\eta|_{^1\vartheta(t)},\,^1\vartheta(t)\in\,J^1M_4\}$. Thus, we can restrict the considerations to the case when $\eta$ is a Lorentzian metric. Although the cosntruction of the average structure for Lorentzian metrics is not universal, we stick to a Lorentzian character for $\eta$ in the following paragraphs, due to the technical simplicity of the theory. 

The Lorentzian metric $\eta$ is also the formal limit of the metric of maximal proper acceleration $g$ when  $A_{\textrm{max}}\to +\infty$ and the operator $\nabla$ is the covariant derivative operator of the Levi-Civita connection of $\eta$.

 The dot-notation, that we will use often,  refers to the derivative with respect to the proper time $\tau$ determined by the metric $\eta$ along $\vartheta$, which is defined by the standard relativistic expression
\begin{align}
\tau[\vartheta]:=\,\int_{\vartheta}dt\,\left(-\eta\left(\frac{d{\vartheta}(t)}{dt},\frac{d{\vartheta}(t)}{dt}\right)\right)^{1/2} .
\label{propre time of eta}
\end{align}

We can now formulate the notion of metric of maximal proper acceleration.
\begin{definicion}
The bilinear form
\begin{align}
g|_{\,^2\vartheta(\tau)} :=\,\left(1-\, \frac{\eta|_{\vartheta(\tau)}(\nabla_{\dot{\vartheta}(\tau)}\dot{\vartheta}(\tau),
\nabla_{\dot{\vartheta}(\tau)}\dot{\vartheta}(\tau)
)}{A^2 _{\textrm{max}}}\right)\eta|_{\vartheta(\tau)} .
\label{maximalaccelerationmetric}
\end{align}
is the metric of maximal proper acceleration.
 The structure $(M_4,g)$, where $M_4$ is a $4$-dimensional manifold and $g$ is the metric defined by the relation \eqref{maximalaccelerationmetric}, is a spacetime of maximal proper acceleration.
 \end{definicion}

The proper acceleration is defined by the expression
\begin{align*}
a^2:=\,\eta(\nabla_{\dot{\vartheta}}\dot{\vartheta},\nabla_{\dot{\vartheta}}\dot{\vartheta}),
\end{align*}
 The metric $g$ determines the physical proper time along $\vartheta$. It is given by the expression
\begin{align}
s [\vartheta] :=\,\int_{\vartheta} \,d\tau\,\left( -g|_{\,^2\vartheta(\tau)}(\dot{\vartheta},\dot{\vartheta})\right)^{1/2} = \,\int_{\vartheta} \,d\tau\,\left(1-\frac{a^2}{A^2_{\textrm{max}}}\right)^{1/2},
\label{propertime g}
\end{align}
where the relation $\eta(\dot{\vartheta},\dot{\vartheta})=\,-1$ has been applied. The value of the maximal proper acceleration is $A_{\textrm{max}}$, in the sense that $a^2 <\,A^2_{\textrm{max}}$. The value of the maximal proper acceleration depends on the interaction considered. For electrodynamic systems and gravitational systems, the scales of the corresponding maximal proper accelerations are very different to each other, being up to $20$ orders of magnitude between each other \cite{Ricardo Nicolini 2018}.

The metric of maximal proper acceleration is a {\it classical structure}, in the sense that it does not involve superpositions of spacetime or a notion of quantum spacetime or a graviton particle.

In our theory, it is $g$ and not $\eta$ the metric structure that determines the physical proper time, an hence, the causal character of curves and world lines. Therefore, we have the following definition,
\begin{definicion}
 A curve $\vartheta :I\to M_4$ is timelike if $g(\dot{\vartheta},\dot{\vartheta})< \, 0$, null if $g(\dot{\vartheta},\dot{\vartheta}) =0$ or spacelike if $ g(\dot{\vartheta},\dot{\vartheta})>0$.
 \label{timelike spacelike null}
 \end{definicion}
Similarly, one can define the causal character of a curve with respect to $\eta$.
For world lines with $a^2<A^2_{\textrm{max}}$,  $\eta$ and $g$ are causally equivalent along the second jet $^2\vartheta$ of the  test particle world line, in the sense that the causal character of a given curve, spacelike, null or timelike is the same with respect to $g$ or with respect to $\eta$. In particular, the light-cone structure is the aggregate of curves $\gamma:J\to M_4$ such that conditions
\begin{align}
g|_{\,^2\vartheta(\tau)}({\gamma}',{\gamma}')=\,0
\label{null cone}
\end{align}
hold good for each of them, where the derivative $\gamma'$  is taken with respect to the parameter $t\in J$ of $\gamma$.
The null condition \eqref{null cone} does not depend upon the probe particle second jet $^2\vartheta:I\to J^2M_4$, except for curves of maximal proper acceleration, where the acceleration of the curve $\vartheta$ is the maximal proper acceleration and one has an additional null condition for $g$ than the standard light-cone $\eta(\dot{\gamma},\dot{\gamma})=0$.
\section{The existence of a superluminal sector}
\subsection{Kinematical and mechanical variables}
Let us represent an ideal, sub-luminal observer as a vector field  $\mathcal{O}\in \Gamma TM_4$ such that $\eta(\mathcal{O},\mathcal{O})=\,-1$. We can perfectly well call such observer sub-luminal. This notion is independent of the system it is observing.  Let us consider a timelike vector field $P:I\to M_4$ along $\vartheta$ such that
\begin{align}
g|_{^2\mathcal{\vartheta}}(P,P)=\,-m^2\,c^4.
\label{on-shell condition}
\end{align}
$P$ is the four-momentum of the particle with world line $\vartheta:I\to M_4$. The parameter $m$ is the inertial mass of the particle, which is the mass measured by an instantaneous inertial observer, with the velocity $\dot{\vartheta}$ at $\vartheta(\tau)$, but with zero proper acceleration. It is assumed that the inertial mass $m$ is constant. In order to define geometrically the energy, Newtonian speed and Newtonian momentum of the particle we follow a methodology analogous to the one in  ref.  \cite{BennTucker, Ricardo 2024, Ricardo 2025}. With respect to the observer ${{\mathcal{O}}}$, there is an unique decomposition of the form
\begin{align}
{P}=\,c\,\mathcal{P}+\,\mathcal{E}\,\mathcal{O},
\label{Decomposition of P}
\end{align}
 where
 \begin{align}
 \begin{cases}
 & \mathcal{E}=-g|_{^2\vartheta}(P,\mathcal{O}),\\
 &\eta(\mathcal{O},\mathcal{O})=\,-1,\\
  & g_{^2\vartheta}(\mathcal{P},\mathcal{O})=\,\eta(\mathcal{P},\mathcal{O})=0.
   \end{cases}\label{Definicion covariante energia}
   \end{align}
   $\mathcal{E}$ is the energy of the particle $\vartheta:I\to M_4$ measured by the observer $\mathcal{O}$; the second relation express the condition that $\mathcal{O}$ is sub-luminal; the third condition is that $\mathcal{P}$ is orthogonal to $\mathcal{O}$ and hence, spacelike.
   The {\it Newtonian velocity} is defined in terms of observables defined by $\mathcal{O}$ by the relation ${\bf v} :=\,\frac{\mathcal{P}}{\mathcal{E}}\,c$, as in the relativistic theory, while the velocity squared is defined by ${\bf v}^2:=\,\eta ({\bf v},{\bf v})$. It follows from \eqref{on-shell condition} the following relation,
   \begin{align}
-m^2\,c^4 =\,-\mathcal{E}^2\left(1-\frac{{\bf v}^2}{c^2}\right)\cdot \left(1-\frac{a^2}{A^2_{\textrm{max}}}\right) .
\label{dispersion relation timelike observer}
\end{align}
It follows that the expression for the {\it energy} $\mathcal{E}$ measured by the observer  ${\mathcal{O}}$ in terms of the Newtonian velocity squared ${\bf v}^2$ and the proper acceleration squared $a^2$ of the world line $\vartheta :I\to M_4$ is given by the expression \cite{Caianiello 1992, Ricardo2015}
\begin{align}
\mathcal{E}(\tau)=\,\pm \frac{1}{\sqrt{\left(1-\frac{a^2(\tau)}{A^2_{\textrm{max}}}\right)\,\left(1-\frac{{\bf v}^2(\tau)}{c^2}\right)}}\,m\,c^2 .
\label{modifiedE}
\end{align}
The kinetic energy of the system is defined to be $\mathcal{K}:=\mathcal{E}-\,m\,c^2$.
In a similar way, the Newtonian momentum $\mathcal{P}$ is given by the expression \cite{Caianiello 1992, Ricardo2015}
\begin{align}
\mathcal{P}(\tau)=\,\frac{1}{\sqrt{\left(1-\frac{a^2(\tau)}{A^2_{\textrm{max}}}\right)\,\left(1-\frac{{\bf v}^2(\tau)}{c^2}\right)}}\,m\,{\bf v}(\tau) .
\label{threemoment}
\end{align}

The dynamical constraints for particles with positive mass $m>0$, is that the energy $\mathcal{E}$ must be positive and the Newtonian momentum $\mathcal{P}$ real. Therefore, according to the relations \eqref{modifiedE} and \eqref{threemoment}, there are two kinematical sectors for particles of positive mass $m>0$ admitting a well-defined energy and Newtonian momentum:
\begin{itemize}
\item  The {\it sub-luminal}/{\it sub-acceleron} sector is characterized by the conditions $m>0$, ${\bf v}^2<\,c^2$ and $a^2<\,A^2_{\textrm{max}}$. This is the direct generalization of the standard causal relativistic sector. Therefore, we can also call it the {\it standard sector}.

\item The  {\it super-luminal}/{\it super-acceleron} or in short, the {\it tachyonic sector}, characterized by the conditions $m>0$, ${\bf v}^2>\,c^2$ and $a^2>\,A^2_{\textrm{max}}$. Thus, particles  pertaining to this sector are super-luminal in the sense that ${\bf v}^2> \,c^2$. The particles whose world line pertain to this sector will be called {\it tachyons}.
\end{itemize}
Therefore, the tachyons considered in this article are {\it ordinary matter}, with positive inertial mass $m>0$.

Since in the tachyonic sector the additional condition $1-a^2/A^2_{\textrm{max}}<\,0$ must also hold good, the causal character of curves of $g$ and $\eta$ are inverted.  If a curve is time-like for $\eta$, it is spacelike with respect to $g$ and viceversa. Due to the condition $(1-{\bf v}^2/c^2)<0$, a world line pertaining to the super-luminal/super-acceleron sector is spacelike with respect to $\eta$. Hence, applying standard kinematical relations for $\eta$, one can show that the proper acceleration $a$ could be either timelike or spacelike with respect to $\eta$, depending on the particular case. But if we stick to the condition $a^2/A^2_{\textrm{max}}>1$, the world line will be timelike with respect to the physical metric $g$.

The conditions ${\bf v}^2 =\,c^2$, $a^2=\,A^2_{\textrm{max}}$ become {\it separation conditions} instead than absolute bounds for the kinematics of bound particles: if the coordinates $a^\mu$, ${\bf v}^\mu$ are continuous, then it is impossible to cross such separation conditions to pass from the standard sector to the tachyonic sector or viceversa.

The existence of the tachyonic sector leads to several general consequences. Let a particle with mass $m>0$ be initially in a state such that $(|{\bf v}_1|^2 >\,c^2,\,a^2_1>\,A^2_{\textrm{max}})$ and assume that suddenly the acceleration is ended. Then by the law of inertia, the system will end in a state $({\bf v}^2 =\,|{\bf v}_1|^2 >\,c^2,\,a^2_2=0)$. But this state is physically forbidden, since it implies an imaginary kinetic energy. Therefore, we need to conclude that the proper acceleration of a tachyon cannot change too abruptly to zero. Indeed, according to the theory, the tachyon needs to be accelerated with a extremely large acceleration such that $a^2>\,A^2_{\textrm{max}}$, since at the classical level, it cannot cross the separation condition $(|{\bf v}_1|^2 =\,c^2,\,a^2_1=\,A^2_{\textrm{max}})$.
\subsection{On the notion of super-luminal observers} A super-luminal observer $\widetilde{\mathcal{O}}$  is defined  by a vector field such that $\eta(\widetilde{\mathcal{O}},\widetilde{\mathcal{O}})=+1$. This notion is independent of the particle it observes. Following a similar procedure as for the case of sub-luminal observers, it is obtained the following relation:
\begin{align}
-m^2\,c^4 =\,\mathcal{E}^2\left(1+\frac{{\bf v}^2}{c^2}\right)\cdot \left(1-\frac{a^2}{A^2_{\textrm{max}}}\right) .
\label{dispersion relation spacelike observer}
\end{align}
It follows that if a particle is in the super-acceleron regime $a^2>\,A^2_{\textrm{max}}$,
\begin{align}
\mathcal{E}(\tau)=\,\pm \frac{1}{\sqrt{\left(\frac{a^2(\tau)}{A^2_{\textrm{max}}}-1\right)\,\left(1+\frac{{\bf v}^2(\tau)}{c^2}\right)}}\,m\,c^2 ,
\label{modifiedEsuperluminalobserver}
\end{align}
valid for particles with positive mass and any speed different from the speed of light, but such that $\eta(\ddot\vartheta,\ddot\vartheta)>\,0$. Note that the acceleration is spacelike in the sense that $\eta(\ddot\vartheta,\ddot\vartheta)>\,0$ is non-trivial. For instance, in the case of an hyperbolic world line in dimension $2$, the condition that $\eta(\dot\vartheta,\dot\vartheta)>\,0$ implies that $\eta(\ddot\vartheta,\ddot\vartheta)<\,0$.

The relation \eqref{modifiedEsuperluminalobserver} is valid for any super-luminal observer respect to $\eta$, independently of the proper acceleration of the observer.

For sub-acceleron test particles, an analogous procedure shows that for a super-luminal observer the energy should be given by the expression
\begin{align}
\mathcal{E}(\tau)=\,\pm \frac{\imath}{\sqrt{\left(1-\,\frac{a^2(\tau)}{A^2_{\textrm{max}}}\right)\,\left(1+\frac{{\bf v}^2(\tau)}{c^2}\right)}}\,m\,c^2 .
\label{modifiedEsuperluminalobserver 2}
\end{align}
 Therefore, super-luminal observers cannot measure consistently the energy of sub-acceleron particles, since the energy measured will be imaginary and hence, it cannot be compared with a standard real energy scale by means of a standard order relation or an standard detector. This situation contrasts with the notion of sub-luminal observer, for which a meaningful definition of energy is given by \eqref{modifiedE} for both, standard particles and tachyons. This makes the notion of super-luminal observer represented by a spacelike vector field with respect to $\eta$ incomplete.

A further consequence of the above theory is the non-equivalence between the notion of observer and the notion of test particle, since a sub-luminal observer can measure both standard particles and tachyon particles energy, while a super-luminal observer can only measure physically the energy of a tachyon particle. Moreover, let us remark that while we the metric $g$ to define spacelike and timelike particles, it is used the metric $\eta$ to characterize the notion of sub-luminal and super-luminal observers.
\section{On the possibility of increasing the kinetic energy of a particle during deceleration}\label{gain of energy during deceleration}
We discuss in this {\it section} a surprising effect due to the peculiarities of the notion of kinetic energy in spacetimes of maximal proper acceleration: the possibility of increasing the kinetic energy simultaneously as the particle is decelerating.
\subsection{The sub-luminal sector}
We consider first the case of a deceleration process in the {\it sub-luminal}/{\it sub-acceleron} sector. By deceleration we mean that for $\tau_1<\tau_2$, the Newtonian speed decreases in the sense that $|{\bf v}_2|<\,|{\bf v}_1|$. According to the definition of energy \eqref{modifiedE}, it is possible that the energy of a particle increases simultaneously as it decelerates. The condition for these phenomena to happen is quite general and does not depend on special conditions on the type of acceleration (like being circular motion, or hyperbolic motion), although the amount of increase depends critically on the scale of the maximal proper acceleration.

 An arbitrary change in the energy of a point particle between $\tau_1$ and $\tau_2$ is of the form
\begin{align}
\frac{\mathcal{E}_2-\mathcal{E}_1}{m\,c^2}=\,\frac{1}{\sqrt{1-\frac{a^2_2}{A^2_{\textrm{max}}}}}\,\frac{1}{\sqrt{1-\frac{|{\bf v}_2|^2}{c^2}}}-\,
\frac{1}{\sqrt{1-\frac{a^2_1}{A^2_{\textrm{max}}}}}\,\frac{1}{\sqrt{1-\frac{|{\bf v}_1|^2}{c^2}}} .
\end{align}
Denoting by $\gamma_{A_{\textrm{max}}}[a_k]:=\,{1}/{\sqrt{1-\frac{a^2_k}{A^2_{max}}}}$ for $k=1,2$, the condition for an increase of the energy of a standard particle with positive mass $m >\,0$ when decelerating is casted in the form
\begin{align}\label{deceleration condition}
\begin{cases}
&\gamma^2_{A_{\textrm{max}}}[a_2]>\, \gamma^2_{A_{\textrm{max}}}[a_1]\,\left(\frac{c^2-\,|{\bf v}_2|^2}{c^2-\,|{\bf v}_1|^2} \right),\\
&|{\bf v}_2|^2\, < |{\bf v}_1|^2 .
\end{cases}
\end{align}
Since $|{\bf v}_2|^2\, < |{\bf v}_1|^2$ and $\gamma_{A_{\textrm{max}}}[a_1]\geq 1$, it holds that $\gamma_{A_{\textrm{max}}}[a_2]>1$, which is the necessary condition for having physically consistent values  of energy and Newtonian momentum in the {\it sub-luminal}/{\it sub-acceleron} sector.

Let us consider several special cases to understand until which extend the condition \eqref{deceleration condition} can be fulfilled. The first case is the trivial situation when the proper acceleration is zero during the whole evolution of the system, $a^2=\,0$. It is direct that in this case $|{\bf v}_2| =\,|{\bf v}_1|$ and hence, there is no increase in the energy of the particle. The second case to consider is hyperbolic motion, where the proper acceleration is constant, $a^2 =\,\mathcal{C}$. In this case, the condition \eqref{deceleration condition} is re-casted as
\begin{align}
\begin{cases}
& 1>\, \left(\frac{c^2-\,|{\bf v}_2|^2}{c^2-\,|{\bf v}_1|^2} \right),\\
&|{\bf v}_2|^2\, < |{\bf v}_1|^2 ,
\end{cases}
\label{Hyperbolic case}
\end{align}
which is impossible to be satisfied in the sub-luminal/sub-acceleron sector. Thus, for hyperbolic motion, the effect is not realized in that sector.
Therefore, in order that the effect could be possible in the sub-luminal/sub-acceleron sector, it is required a non-constant proper acceleration.  In this case, the condition for increasing the energy and simultaneously decreasing the speed reads
\begin{align}
\frac{A^2_{\textrm{max}}-\,a^2_1}{A^2_{\textrm{max}}-\,a^2_2} >\, \frac{c^2-\,|{\bf v}_2|^2}{c^2-\,|{\bf v}_1|^2}\,> 1 .
\label{deceleration condition 2}
\end{align}
In particular, the condition \eqref{deceleration condition 2} implies that for an increase of the kinetic energy while decelerating to happen, the proper acceleration of the particle during the process must increase, $a^2_2> a^2_1$.

\subsection{The tachyonic sector} In the super-luminal/super-acceleron sector, the analysis follows similar lines as for the sub-luminal/sub-acceleron sector and leads to analogous results. Thus, the necessary condition for the effect is also the relations \eqref{deceleration condition}. In general, in the tachyonic sector the acceleration $a$ could be either timelike or spacelike with respect to $\eta$. When the proper acceleration is spacelike, the analysis is very similar to the one as for the standard sector.
\section{A toy model illustrating the effect and its amplification}\label{toy model}
The following construction illustrates the increase of the kinetic energy of a point charge particle in a particular scheme of linear deceleration. Let us consider a point particle with charge $q^-<\,0$ approaching an infinitely long wire endowed with uniform charge density $\lambda^- <\,0$. Located far from the wire, the initial acceleration of the particle is assumed to be zero. For small accelerations, the dynamics of the particle can be approximately described by the Lorentz force equation and since the initial velocity is prepared to be directed orthogonally towards the wire, there is an increasing repulsion between the charge and the wire that makes the particle to decelerate. Since the charge decelerates in a straight direction, the radiation emitted is much less than when bending and indeed, we will show the existence of a domain (far from the turning points) where the emission of Larmor's radiation is negligible. Furthermore, we do not consider bremsstrahlung effects.

For the system under consideration, the acceleration of the point particle is not constant, since it has a dependence of the form $1/r(\tau)$, where $r(\tau)$ is the distance of particle to the wire at the instant $\tau$. Therefore, the proper acceleration increases as long as the particle approaches the wire. We can apply the theory described in the previous section, with a deceleration due to the electric field which is continuously increasing during the evolution to show under which conditions the particle could gain energy even if it is decelerating under the interaction with the external field generated of the wire. The gain of kinetic energy the particle by this effect is small, due to the large value of the maximal acceleration in electrodynamics, except if the particles can approach very closely to the wire, where our analysis not applies. Eventually, the particle will stop due to the Coulomb repulsion force and start to return towards the initial position. Another restriction in the analysis avoids situations when
the particle has a large initial speed, it can reach closer to the wire and due to the nature of the electromagnetic field \cite{Ricardo 2025b} and quantum mechanical effects \cite{Sauter, Schwinger}, it could suffer very intense interactions. We assume that the initial conditions are such that we are far from that regime.

 As discussed in Appendix \ref{Appendix B}, the model proposed for the motion of a point particle is formally the Lorentz-Dirac equation without the Schott's term in a spacetime of maximal proper acceleration. In detail, the world lines of a charged point particle are solutions of the differential equation
  \begin{align}
m_0\,\ddot{\vartheta}^{\mu} =\,q\,F^{\mu}\,_{\nu}\,\dot{\vartheta}^{\nu}-
\,\frac{2}{3}\,{q^2}\,\ddot{\vartheta}^{\nu}\ddot{\vartheta}_{\nu}\,\dot{\vartheta}^{\mu},\quad F^{\mu}\,_{\nu}=\,g^{\mu\rho}F_{\rho\nu}, \quad \mu,\rho,\nu =0,1,2,3,
\label{equationofmotionnoncovariant expression}
\end{align}
where $g_{\mu\rho}$ is the metric of maximal proper acceleration and the external electromagnetic field $F_{\rho\nu}$ is the external electromagnetic field in the limit $A_{\textrm{max}}\to +\infty$. Note that for a higher order derivative Lagrangian as in the case of the Lagrangian of a point charged particle described in Appendix \ref{Appendix B}, the equations of motion is of fourth order. However, the equation \eqref{equationofmotionnoncovariant expression} is not directly the extremal condition for the Lagrangian \eqref{Lagrangian}, but it is obtained under phenomenological restrictions, in particular, after imposing to the Euler-Lagrange equations that the equation must be of second order and compatible with the relativistic Larmor's radiation law.

Let us remark that the kinetic condition for the particle is that $g(\dot\vartheta,\dot\vartheta)=-1$ and not necessarily $\eta(\dot\vartheta,\dot\vartheta)=-1$, hence avoiding one of the common troubles with eliminating the Schott's term as result of the usual orthogonality between $a$ and the proper velocity $\dot{\vartheta}$.

Since we are considering small accelerations in the sense that $\epsilon=\eta(a,a)/A^2_{\textrm{max}}<< 1$, the term $\frac{2}{3}\,{q^2}\,\ddot{\vartheta}^{\nu}\ddot{\vartheta}_{\nu}\,\dot{\vartheta}^{\mu} =\,\frac{3}{2} \,\epsilon\,\beta\,A_{\textrm{max}}$, where $\beta^2=\,\frac{{\bf v}^2}{c^2}$ is small compared with the Lorentz force $q\,F^{\mu}\,_{\nu}\,\dot{\vartheta}^{\nu}$. If such a term is formally treated as a perturbation of the Lorentz Force equation, then the corresponding velocity is such that $\dot\vartheta^\mu =\dot\vartheta^\mu_L+\,\mathcal{O}(\epsilon\,\beta\, A_{\textrm{max}})$, where $\dot{\vartheta}_L$ is the velocity vector of the world line solution of the Lorentz force equation. Inserting this velocity on the expression \eqref{modifiedE} for the kinetic energy and under the assumption $\beta>> \mathcal{O}(\epsilon A_{max}\,\beta)$, although it is of the same order than $\sqrt{\epsilon}$, it can be shown that the kinetic energy is approximately equal as if one used the solution of the Lorentz force equation.

By the above argument we can make the approximation that the dynamics of the particles is described by the Lorentz force equation. This approximation breaks when $a^2\sim A^2_{\textrm{max}}$, among other circumstances. For a point particle in an external field generated by a linear infinite charge line, the Lorentz force equation is of the form
\begin{align}
m\,\frac{d^2 r}{dt^2} =\,\frac{q\,\lambda}{2\pi \epsilon_0\,r}.
\label{Lorentz force for the example}
\end{align}
It can be re-cast as
\begin{align}
m\,\frac{d{\bf v}}{dr}\,{\bf v}=\, \frac{q\,\lambda}{2\pi \epsilon_0\,r} .
\label{segunda forma de poner la ecuación de Lorentz}
\end{align}
It is straightforward to integrate this equation. For the admissible initial position $r=L_0$ and initial speed is ${\bf v}_0$, the solution is
\begin{align}
\frac{1}{2}\,m\,\left({\bf v}^2-\,{\bf v}^2_0\right)\approx \,\frac{q\,\lambda}{2\pi\,\epsilon_0}\,\ln\left(\frac{r}{L_0}\right).
\end{align}
This energy conservation condition is the condition \eqref{balance of energy in a spacetime of maximal acceleration} of Appendix \ref{Appendix C} but subjected to the approximation that the velocity ${\bf v}$ corresponds to a solution of the Lorentz force equation, that is, the condition \eqref{approximate balance equation}. As such, the condition is only approximately valid condition, which is useful to us only to approximately determine the value of the velocity. Indeed, according to the theory developed in \ref{Appendix C}, the appropriate balance equation is of the form
\begin{align}
\frac{1}{2}\,m\,\left({\bf v}^2-\,{\bf v}^2_0\right)+ \frac{1}{2}\,m\,c^2\left(\frac{a^2}{A_{\textrm{max}}}-\,\frac{a^2_0}{A^2_{\textrm{max}}}\right)\,\approx \frac{q\,\lambda}{2\pi\,\epsilon_0}\,\ln\left(\frac{r}{L_0}\right).
\end{align}
Note that if approximations to evaluate ${\bf v}$ and $a$ are assumed, as we did by solving the Lorentz force equation, then the balance equations will be formally un-satisfied. Such a discrepancy is only an artifact of the approximations.

The process of increasing the energy finishes either by incidental reasons such as when the particle reaches to close to the charged wires, producing at the end a bremsstrahlung process or if significant radiative losses mitigates the gain due to the deceleration. We consider here when radiation losses can be neglected. Applying Larmor's power law for linear acceleration we obtain the relation \cite{Jackson},
\begin{align}
P_{rad} =\,\frac{2}{3}\, \frac{q^2}{m^2\,c^3}\cdot\,\left(\frac{d|{\bf p}|}{dt}\right)^2 =\,\frac{2}{3}\, \frac{q^2}{m^2\,c^3}\cdot\,\left(\frac{d\mathcal{E}}{dr}\right)^2 .
\label{linear Larmor law}
\end{align}
On the other hand, the power provided by the external field is
\begin{align*}
P=\,\frac{d\mathcal{E}}{dt}=\,\frac{d\mathcal{E}}{dr}|{\bf v}| .
\end{align*}
 The requirement that the energy loss by radiation is small compared with the energy gain rate of the particle leads to the condition
 \begin{align}
|{\bf v}| \gg \,\frac{2}{3}\, \frac{q^2}{m^2\,c^3}\cdot\,\left(\frac{d\mathcal{E}}{dr}\right).
 \label{disregard power}
 \end{align}
 This condition excludes the turning points where $|{\bf v}|=0$ from the domain when radiation emission can be neglected and therefore, also from our current considerations.

 In order to illustrate the effect of increasing energy while decelerating, let us assume  that the condition \eqref{disregard power} holds good.
 It follows from the definition of the energy \eqref{modifiedE} and the Lorentz force equation written in the form \eqref{segunda forma de poner la ecuación de Lorentz}  that
 \begin{align}
 \frac{d\mathcal{E}}{dr}\approx \frac{q\lambda}{2\,\pi\epsilon_0\,r}\,\left(1-\frac{q\lambda}{2\,\pi\epsilon_0\,r}\,\frac{c^2}{A^2_{\textrm{max}}}\frac{1}{mr}\right) .
 \end{align}
 Inserting the leading order term of this expression in the condition \eqref{disregard power} and inserting the value of the velocity as a function of the position, leads to the constraint,
 \begin{align}
 |{\bf v}_0|^2+\frac{q\lambda}{\pi\,\epsilon_0 \, m}\,\ln \left(\frac{r}{L_0}\right)\gg\,\left(\frac{2}{3}\, \frac{q^2}{m^2\,c^3}\cdot\,\left(\frac{d\mathcal{E}}{dr}\right)\right)^2\approx \,\left(\frac{2}{3}\, \frac{q^2}{m^2\,c^3}\cdot\,\left(\frac{q\,\lambda}{2\pi\,\epsilon_0 \,r}\right)\right)^2
 \end{align}
 One domain where this constraint holds good is around the initial position $r\approx L_0$ and if the initial speed is large enough (but still small compared with the speed of light). In such a situation, the system needs to satisfy the condition
 \begin{align}
 |{\bf v}_0|\,\gg \frac{2}{3}\, \frac{q^2}{m^2\,c^3}\cdot\,\left(\frac{q\,\lambda}{2\pi\,\epsilon_0 \,L_0}\right).
 \label{constrain on v and L}
 \end{align}

 Now we check that the condition \eqref{deceleration condition 2} holds good in the proximity of the initial location. When applying the Lorentz force equation, condition \eqref{deceleration condition 2} is re-casted as
 \begin{align*}
 \frac{A^2_{\textrm{max}}-\left(\frac{q\lambda}{2\pi\,\epsilon_0L_0 \,m}\right)^2}{A^2_{\textrm{max}}-\left(\frac{q\lambda}{2\pi\,\epsilon_0 r \,m}\right)^2}>\frac{c^2-\left( |{\bf v}_0|^2+\frac{q\lambda}{\pi\,\epsilon_0 \, m}\,\ln \left(\frac{r}{L_0}\right)\right)}{c^2-{\bf v}^2} .
 \end{align*}
 After some algebra, this condition reduces to
 \begin{align*}
 \frac{\left(\frac{q\lambda}{\pi\,\epsilon_0 \, m}\right)^2\,\left(\frac{1}{L^2_0}-\,\frac{1}{r^2}\right)}{1-\frac{\left(\frac{q\lambda}{\pi\,\epsilon_0 \, m}\right)^2\frac{1}{r^2}}{A^2_{\textrm{max}}}} <\,\frac{\frac{q\lambda}{\pi\,\epsilon_0 \, m}\,\ln \left(\frac{r}{L_0}\right)}{c^2-{\bf v}^2}
 \end{align*}
When $r=L_0-\delta$ with $\delta$ small but positive, indicating that the particle is approaching the charged wire, the logarithmic function is $\ln(r/L_0)\approx\,-\delta/L_0$. On the other hand the function $\frac{1}{L^2_0}-\,\frac{1}{r^2}\approx 1/L^2_0 (-2\delta/L_0)$. Since $1-\frac{\left(\frac{q\lambda}{\pi\,\epsilon_0 \, m}\right)^2\frac{1}{r^2}}{A^2_{\textrm{max}}}>0$ and being all other factors positive constants, comparing $-2\delta$ with $-\delta$ one concludes that for $\delta$ small enough, the condition \eqref{deceleration condition 2} holds good.

\section{Conclusion and discussion}
In this article we have discussed the notion of kinetic energy in a spacetime of maximal acceleration and two of its surprising consequences: the existence of a superlumninal sector for positive mass particles and the possibility to increase the kinetic energy of a particle at the same time that its velocity decreases.

In our theory, the mass of a point particle is real and positive $m$ in the sense of the definition \eqref{on-shell condition}, but for particles with an acceleration larger than $A_{\textrm{max}}$. This also applies to particles pertaining to the tachyonic sector. This contrasts with other schemes and theories allowing the existence of tachyons where the mass of the tachyon is imaginary \cite{Recami, Mantegna M. 2015, Castro-Perelman 2025, Castro-Perelman 2025b}. Thus, although in our theory the conditions for tachyons are very difficult to achieve, because they require for the tachyons to be tremendously accelerated beyond the {\it critical acceleration} $A_{\textrm{max}}$, it is conceivable that matter exists already in that sector. On the other hand, note that in our definition of the tachyonic sector, a world line of a test particle is timelike with respect to $g$, implying that for each of the two sectors discussed, the world lines of test particles are timelike with respect to the physical metric structure $g$, generalizing the standard causal condition for massive particles in Lorentzian spacetimes.

We have also concluded the impossibility for a super-luminal observer to measure the energy of a standard particle, since the eventual energy measured will be imaginary and a imaginary energy cannot be transferred to a real energy in the form of a work process by a projection operation or device.

We would like to express here a wild speculation. We have noticed in this work that the speed of light and the maximal acceleration could constitute the boundary of two different kinematical regimens rather than absolute limits. In the classical framework discussed, they are separate regimens, isolated from each other by the fact that passing from one to the other implies to pass by a divergence $|{\bf v}|=c$, $a^2 =\,A^2_{\textrm{max}}$ of the kinetic energy. But at a quantum level, this could be different. If there exists an underlying discrete structure of the observables, then crossing the separation between the two regimens does not require to pass by the kinetic divergence. In the context of our theory this is very intriguing, since as we have seen, tachyons in our theory are standard matter. Thus the potential possibility, in a theory with discrete spacetime and maximal acceleration, of sending superluminal signals by using ordinary matter.

The second result that we have discussed in this article is that, in a spacetime of maximal proper acceleration, the deceleration of a point particle, in the sense of a decrease of the Newtonian velocity, can be accompanied by an increase in its kinetic energy. This effect is consequence of the expression \eqref{modifiedE} for the energy of a point particle in the theory of spacetimes of maximal proper acceleration. Although paradoxical, the increase of the kinetic energy during deceleration discussed in our theory does not imply the violation of the energy-momentum conservation law. The kinetic energy gained  by the particle from the external field is shared into two contributions to the kinetic energy of the particle: one associated to the speed of the particle and a second one associated to the proper acceleration of the particle. These two contributions to the energy are respectively encoded by the relativistic gamma factor $\gamma[{\bf v}(\tau)]$ and the maximal accelerator factor $\gamma_{A_{\textrm{max}}}[a(\tau)]$ of the particle. While the contribution due to the velocity factor decreases during the deceleration, the contribution due to the acceleration factor can increase in such a way that the balance between both contributions to the change of kinetic energy can be positive. This is the physical interpretation of the condition \eqref{deceleration condition 2}.

The above considerations suggest the equivalence between acceleration and velocity as contributors to the kinetic energy. This equivalence may be interpreted in the form of a {\it reciprocity  principle}, where velocity and acceleration are on equal footing from an energetic viewpoint.  Let us remark the compatibility of such reciprocity principle in jet spacetime with general covariance. Indeed, the expressions for the  energy and Newtonian momentum  \eqref{modifiedE} and \eqref{threemoment} are formulated by means of covariant objects and operations.

There is certainly a similarity between the proposed reciprocity or equivalence of velocity and acceleration from the point of view of energy described above and the reciprocity principle proposed by Born \cite{Born 1938, Born 1949}, namely, the principle by which coordinate and momentum variables should be treated on equal footing in fundamental physical description.
Modern versions and generalizations of Born's reciprocity principle are formulated in terms of phase space variables and a symplectic-group theoretic point of view \cite{Low1993,Low 2002,Castro 2005a,Castro 2005b}, often in terms of representation theory of a flat metric in phase space.  Using coordinates $(x^\mu, p^\mu)$ in a covariante $8$-dimensional phase space, the metric {\it Born-Green metric} is
\begin{align}
 d s^2 =\,\frac{1}{c^2}\left(-c^2\,(dx^0)^2+(d\vec{x})^2\right)+\,\frac{1}{b^2}\left(-\,(\,p^0)^2 +c^2(d\vec{p})^2\right) ,
 \label{Born-Green metric}
 \end{align}
where $b$ is a parameter representing the maximal force. Reciprocity relativity is partially driven by studying the representation theory and the consequences of postulating this metric structure as the structure defining the physical proper time of a point particle.

In comparison, the theory discussed in the present work suggests an energetic equivalence between velocity and acceleration and is developed in terms of  jet geometry associated to a classical system.
Indeed, {\it the high order jet geometry reciprocity principle} can be formulated geometrically by considering the existence of a $2$-form that reproduces infinitesimally the physical proper time \eqref{propertime g}. A $2$-form with the properties of a metric and consistent with the principle of equivalence between spacetime coordinates and velocity coordinates transpositions is the Sasaki-type metric \cite{Ricardo Paolo Piccione 2015}, that in local  $1$-forms $(d x^\mu,\delta {\bf v}^\mu)$ for flat spacetimes reads
\begin{align}
 d s^2 =\,\frac{1}{c^2}\left(-c^2\,(dx^0)^2+(d\vec{x})^2\right)+\,\frac{1}{A^2_{\textrm{max}}}\left(-\,(\,\delta{\bf v}^0)^2 +(\delta\vec{{\bf v}})^2\right) ,
 \label{Caianiello metric}
 \end{align}
 where the $1$-forms $\delta {\bf v}^\mu$ are constructed using the non-linear connection of $\eta$ \cite{GallegoPiccioneVitorio:2012, Ricardo 2021, Ricardo 2024c}.
 The metric \eqref{Caianiello metric} is the covariant version of Caianiello's quantum geometry metric \cite{Ricardo2007}. It reproduces the physical proper time associated to the metric of maximal proper acceleration \eqref{Hyperbolic case} when acting on timelike and spacelike vectors. The analogy between the metrics \eqref{Born-Green metric} and \eqref{Caianiello metric} is clear, preluding a close relation between both theories.

The effect of increasing the kinetic energy during deceleration has been illustrated by the system composed by a point charged particle interacting with an infinite long charged line. This example has been developed in the non-relativistic regime, since the contributions associated to the acceleration $a$ and the velocity $|{\bf v}$ are of the same formal order. The treatment of the example is highly idealized, intended to illustrate the general theory of {\it section} \ref{gain of energy during deceleration}. Indeed, our discussion disregards radiation losses and bremsstrahlung effects and makes the approximation for particle world lines as solutions of the Lorentz force equation, instead than from the equation of motion \eqref{equationofmotionnoncovariant expression}. It also assumes idealizations for the external sources of fields as being infinitely large charge distributions, disregards {\it border effects} and overlooks the details for the mechanism to provide to the particle the desired initial conditions. Therefore, although the effect is illustrated by the example, it is very desirable to have a more accurate description of the system overcoming some of such limitations.

\appendix
\section{Higher order jet Electrodynamics}\label{Appendix A}
In this {\it Appendix} we provide a short introduction to higher order electrodynamics. Since this topic has  been considered elsewhere with relative extension \cite{Ricardo2012,Ricardo 2024}, we discuss here only the elements necessary to introduce the action functional for a point particle in an external field and to discuss the stress-energy tensor and the energy balance conditions.
\subsection{Generalized higher order jet electromagnetism}
Let $M_4$ be a smooth $4$-manifold.
\begin{definicion}
The electromagnetic field $\bar{F}$ along the lift $j^k x:I\to J^k_0(M_4)$ is a $2$-form  that in local natural coordinates can be written as
\begin{equation}
\bar{F}(j^kx)=\bar{F}(x,{x}',{x}'',{x}''',..., x^{(k)})=\big(F_{\mu\nu}(x)+
\Upsilon_{\mu \nu}(x,{x}',{x}'',{x}''',..., x^{(k)})\big)d_4 x^{\mu}\wedge d_4 x^{\nu}.
\label{electromagneticfield}
\end{equation}

The excitation tensor $\bar{G}$ along the lift $j^kx:I\to J^k_0(M_4)$ is a $2$-form
\begin{equation}
\bar{G}(j^kx)=\bar{G}(x,{x}',{x}'',{x}''',..., x^{(k)})=\big(G_{\mu\nu}(x)+
\Xi_{\mu \nu}(x,{x}',{x}'',{x}''',..., x^{(k)})\big)d_4 x^{\mu}\wedge d_4 x^{\nu}.
\label{excitationtensor}
\end{equation}

The density current $\bar{J}$ is  a $3$-form
\begin{equation}
\bar{J}(x,{x}',{x}'',{x}''',..., x^{(k)})=\,\big(J_{\mu\nu\rho}(x)\,+
\Phi_{\mu\nu\rho}(x,{x}',{x}'',{x}''',..., x^{(k)})\big)\,d_4x^{\mu}\wedge d_4x^{\nu}\wedge d_4x^{\rho}.
\label{generalizedcurrent}
\end{equation}
\end{definicion}
To each of these three types of generalized fields there are associated standard fields $F,G\in \,\Lambda^2 M_4$ and $J\in\,\Lambda^3M_4$ defined locally by the components $F_{\mu\nu}(x)$, $G_{\mu\nu}(x)$ and $J_{\mu\nu}(x)$ respectively.

The generalized homogeneous of Maxwell's equations are of the form
\begin{align}
d_4\bar{F}=0.
\label{homogeneousequation}
\end{align}
This equation allows the introduction of $1$-forms $\bar{A}$ such that $d_4\bar{A}=\, \bar{F}$ and with structure $\bar{A}(\,^2x)= A(x)+\Omega (\,^kx)$, being $A$ a standard $1$-form over the spacetime manifold $M_4$.

If we assume the constitutive relation $\bar{G}=\,\star \bar{F}$, then the generalized inhomogeneous Maxwell's equations are
\begin{align}
d_4\star \,\bar{F}=\,{J}+d_4\,\star \Upsilon.
\label{nonhomogeneous equation}
\end{align}

From equations \eqref{homogeneousequation} and \eqref{nonhomogeneous equation}
it is possible to construct an effective theory which is equivalent to the standard Maxwell's theory. In particular, the Faraday tensor $F$ must be such that
\begin{align}
d F=0
\label{equationforF}
\end{align}
holds good; $G$ must be a solution of the equation
\begin{align}
d\star\, F=\,J
\label{equationfor*F}
\end{align}
and the conservation of the current density is
\begin{align}
d{J}=0.
\label{equationforJ}
\end{align}
\begin{comentario}
Let us make a remark on the notion of {\it external field}. By external field we mean that the test particle is immersed in a field that excludes its own field. The external field, on the other hand, depends upon the particle's $k$-jet. This asymmetry can be seen as an approximation in situations where the external field is dominant, although effects due to maximal proper acceleration are still noticeable.
\end{comentario}

The theory of higher order jet electrodynamics, together with the theory of spacetimes of maximal proper acceleration was applied in \cite{Ricardo2012, Ricardo 2017} to deduce a second order equation of motion for the classical point particle from geometric considerations and physical phenomenological constraints. The approach followed was based on Rohrlich's derivation of the Lorentz-Dirac equation \cite{Rohrlich}, following a combination of physical insights and jet theory methods. According to this theory, the equation of motion of a point charged particle is of the form
\begin{align}
 m\,\ddot{\vartheta}_{\mu} =\,q\,F_{\mu\nu}\,\dot{\vartheta}_{\nu}-
\,\frac{2}{3}\,{q^2}\,\ddot{\vartheta}^{\nu}\ddot{\vartheta}_{\nu}\,\dot{\vartheta}_{\mu},\quad \mu,\rho,\nu =0,1,2,3,
\label{equation of motion}
\end{align}
which is equivalent to the equation \eqref{equationofmotionnoncovariant expression}. Note that in this equations, the dot-derivatives are taken with respect to the proper time of the metric $\eta$. Since in the theory of spacetimes of maximal proper acceleration, the standard kinematical relation $\eta(\dot\vartheta, \dot\vartheta)=-1$ does not hold. Instead, one has that $g(\vartheta',\vartheta')=-1$ holds good, where the derivatives are with respect to the proper parameter of $g$ \cite{Ricardo2012, Ricardo 2017, Ricardo 2024}. This makes a substantial different with other theories of classical electrodynamics, like Bonnor's theory \cite{Bonnor}, that follow the standard relativistic kinematics.

\section{The action functional of a charged point particle and the equation of motion}\label{Appendix B}
We discuss in this Appendix a reformulation of the model of a point charged particle in an external field in terms of an action functional. This more physical approach than the one discussed previously in \cite{Ricardo2012,Ricardo 2017} has the advantage of providing the background for the derivation of a natural stress-energy tensor, from where conservations laws can be derived.

Let us consider a generalized external field such that $\bar{F}_{ext}=\,d_4 \bar{A}$. The action functional of a point test particle with world line $\vartheta:I\to M_4$ on the external field in a spacetime of maximal proper acceleration is
\begin{align}
S[\,^k\vartheta,g|_{\,^2\vartheta}]=\,\int_I d\tau\left( m_b\,c \,\left(-g|_{\,^2\vartheta}(\dot{\vartheta},\dot{\vartheta}\right)^{1/2}+q\,\dot{\vartheta}^\mu\bar{A}_{\mu} \right) .
\label{action funcional}
\end{align}
where here $m_b$ is the {\it bare mass}.
Therefore, the corresponding Lagrangian is of the form
\begin{align}
L\left(\,^2\vartheta,g|_{\,^2\vartheta} \right)=\, m_b\,c \,\left(-g|_{\,^2\vartheta}(\dot{\vartheta},\dot{\vartheta}\right)^{1/2}+q\,\dot{\vartheta}^\mu\bar{A}_{\mu} .
\label{Lagrangian}
\end{align}
Being a Lagrangian depending upon second derivatives in the variables of $\vartheta^\mu$, the extremal conditions are solutions of the equations \cite{Nesterenko 1989, Lovelock Rund}
\begin{align}
\frac{\partial L}{\partial \vartheta^\mu}-\,\frac{d}{dt}\,\frac{\partial L}{\partial \dot{\vartheta}^\mu}+\,\frac{d^2}{dt^2}\,\frac{\partial L}{\partial \ddot{\vartheta}^\mu}=0,\,\quad \mu =0,1,2,3 .
\label{Lagrange equations for point particle}
\end{align}
Let us expand the Euler-Lagrange equations in the form
\begin{align}
\frac{\partial L}{\partial \vartheta^\mu}\,-\frac{\partial^2 L}{\partial \dot{\vartheta}^\mu\partial \vartheta^\nu}\,\dot{\vartheta}^{\nu}\,-\frac{\partial^2 L}{\partial \dot{\vartheta}^\mu\partial \dot{\vartheta}^\nu}\,\ddot{\vartheta}^{\nu}+\frac{d^2}{dt^2}\,\frac{\partial L}{\partial \ddot{\vartheta}^\mu}=0,\,\quad \mu =0,1,2,3 .
\label{extended equation of Euler-Lagrange}
\end{align}
For a generic Lagrangian $L$ depending up to second order derivatives, the Euler-Lagrange equations will be at least a system of fourth order differential equations.
However, from the experience gained from the study of the Lorentz-Dirac equation for a point charged particle, to have higher order derivatives leads to un-physical solutions. Therefore, we develop a procedure to reduce the order of the equation up to a second order order equation.
First, we make the assumption that the last term in the equation \eqref{extended equation of Euler-Lagrange} plus other higher order terms that could appear in the equations containing higher derivative than $2$ of $\vartheta^\mu$ vanish:
\begin{align}
\frac{d^2}{dt^2}\,\frac{\partial L}{\partial \ddot{\vartheta}^\mu}+\,\textrm{other higher order terms}=0 .
\label{constraint}
\end{align}
Second, we introduce now the renormalization of the mass,
\begin{align}
m_b\rightarrow \, m=\,(1-\epsilon)^{1/2}\,m_b .
\label{renormalization of the mass}
\end{align}
Third, we make the following ansatz for the generalized vector potential $1$-form. Being $\bar{A}=\,A+\Omega$, we assume that $\Omega$ has locally the structure
\begin{align}
\Omega_\lambda =\,f_2(\vartheta)\,\ddot{\vartheta}_\lambda \eta(\dot{\vartheta},\dot{\vartheta})+f_3(\vartheta) \dddot{\vartheta}_\lambda+...
\end{align}
This ansatz for the potential $\bar{A}$ implies that most of the higher order derivatives in the Euler-Lagrange equations comes from the term $\frac{d^2}{dt^2}\,\frac{\partial L}{\partial \ddot{\vartheta}^\mu}$, from the term proportional to $f_3$ in $\bar{A}$ through the derivative $\frac{\partial^2 L}{\partial \dot{\vartheta}^\mu\partial {\vartheta}^\nu}\,\dot{\vartheta}^{\nu}$ and from the terms proportional to $f_k$, with $k\geq 4$. We will see below that also the term proportional to $f_2$ contains a third order term too.
Hence, in order to partially ensure the absence of third order derivatives terms in the Euler-Lagrange equations and in the Lagrangian itself, the condition $f_3=0$ is imposed.
More generally, the imposition of the constraint \eqref{constraint} leads to constraints on the functions $f_3$, $f_4$..., that are otherwise arbitrary: $f_k$ are set up equal to zero for $k>4$ too.
The remaining term in the Lagrangian corresponding to the $\Omega$-piece of the potential is of the form
\begin{align*}
L_\Omega:=\,q\,f_2(\vartheta)\,\left(\ddot{\vartheta}_\lambda\,\dot{\vartheta}^{\lambda}\right)\,\eta(\dot{\vartheta},\dot{\vartheta}) .
\end{align*}
Let us assume that $f_2$ is constant. This condition is rather useful. First, it ensures that each of the terms in $\frac{d^2}{dt^2}\,\frac{\partial L}{\partial \ddot{\vartheta}^\mu}$, specifically, the ones coming from $\Omega$, are proportional to higher order derivatives of the coordinates $\vartheta^\mu$. Second, the remaining contributions to the Euler-Lagrange equation are of the form
\begin{align*}
\begin{cases}
&\frac{\partial q\,\Omega_\lambda\dot{\vartheta}^\lambda}{\partial \vartheta^\mu}=\,0,\\
&\frac{\partial^2 q\,\Omega_\lambda\dot{\vartheta}^\lambda}{\partial \dot{\vartheta}^\mu\partial \vartheta^\nu}\,\dot{\vartheta}^{\nu}=\,0\\
&\frac{\partial^2 q\,\Omega_\lambda\dot{\vartheta}^\lambda}{\partial \dot{\vartheta}^\mu\partial \dot{\vartheta}^\nu}\,\ddot{\vartheta}^{\nu}=\,3\,f_2\left(\dot{\vartheta}_\nu\,\ddot{\vartheta}^\nu\right)\,\ddot{\vartheta}_\mu+\,2\,f_2\,
\left(\ddot{\vartheta}_\nu\,\ddot{\vartheta}^\nu\right)\,\dot{\vartheta}_\mu .
\end{cases}
\end{align*}
The term $3\,f_2\left(\dot{\vartheta}_\nu\,\ddot{\vartheta}^\nu\right)\,\ddot{\vartheta}_\mu$ is of third order, since
$g(\dot{\vartheta},\dot{\vartheta})=-1$ implies $d/d\tau (g(\dot{\vartheta},\dot{\vartheta}))=0$, or
\begin{align*}
\frac{d}{d\tau} (g(\dot{\vartheta},\dot{\vartheta}))=\,\frac{d}{d\tau} \left(\left(1-\frac{a^2}{A^2_{\textrm{max}}}\right)\,\eta(\dot{\vartheta},\dot{\vartheta})\right) =\,\dot{\epsilon}\,\eta(\dot{\vartheta},\dot{\vartheta})+\,\left(1-\frac{a^2}{A^2_{\textrm{max}}}\right)\,\eta(\dot{\vartheta},\ddot{\vartheta}) =0 ,
\end{align*}
where $\epsilon=\,\eta(\ddot{\vartheta},\ddot{\vartheta})/A^2_{\textrm{max}}$. This expression leads to the relation
\begin{align*}
\eta(\dot{\vartheta},\ddot{\vartheta})=\,-\,\frac{\dot{\epsilon}}{2(1-\epsilon)}\eta(\dot{\vartheta},\dot{\vartheta}),
\end{align*}
showing that the term in the Euler-Lagrange equations corresponding  $\Omega$-term in the potential contains a piece with third order derivatives. This terms must also included in the constraint \eqref{constraint}.

Finally, we impose the condition of compatibility with the relativistic Larmor's law, constraining the value of $f_2$. This last constraint ensures conservation of energy-momentum in radiation processes, by assuming that Larmor's law is exact in high order electrodynamics. Larmor's law is well established experimentally and by assuming it in higher order electrodynamics, we are lead to a second order differential equation model for the point charged particle whose validity is linked to Larmor's law validity. The second term is made compatible with the Larmor law if $f_2= \,-\frac{2}{3}\,q $.

In a nut-shell, the model of the point particle is determined by the following three relations:
\begin{align}
\begin{cases}
& m\,\ddot{\vartheta}_{\mu} =\,q\,F_{\mu\nu}\,\dot{\vartheta}_{\nu}-
\,\frac{2}{3}\,{q^2}\,\ddot{\vartheta}^{\nu}\ddot{\vartheta}_{\nu}\,\dot{\vartheta}_{\mu},\quad \mu,\nu =0,1,2,3,\\
& \frac{d^2}{dt^2}\,\frac{\partial L}{\partial \ddot{\vartheta}^\mu}+\frac{3}{2}\,\frac{q}{m}\,\left(\frac{\dot{\epsilon}}{1-\epsilon}\,\eta(\dot{\vartheta},\ddot{\vartheta})\right)
\,\eta(\dot{\vartheta},\ddot{\vartheta})=0, \\
& m =\,{m}_b{(1-\epsilon)^{1/2}} .
\end{cases}
\label{equation of motion}
\end{align}
The bare mass parameter $m_b$ is such that $m$ is constant. 

Several comments are in order.
\begin{itemize}
\item If $m$ is the inertial mass of the particle, then it coincides with the mass when the particle does not have acceleration, while $m_b$ coincides with the mass of the particle when it experiences acceleration. This law is the modification of the inertia due to acceleration discussed in \cite{Ricardo 2024}.

\item Let us note that
\begin{align*}
m_b\,c \,\left(-g|_{\,^2\vartheta}(\dot{\vartheta},\dot{\vartheta})\right)^{1/2} =m\,c \,\left(-\eta(\dot{\vartheta},\dot{\vartheta})\right)^{1/2} .
\end{align*}
Hence the action functional \eqref{action funcional} is equivalent to
\begin{align}
S[\,^k\vartheta,\eta]=\,\int_I d\tau\left( m\,c \,\left(-\eta(\dot{\vartheta},\dot{\vartheta})\right)^{1/2}+q\,\dot{\vartheta}^\mu\bar{A}_{\mu} \right) .
\label{action funcional 2}
\end{align}
Note that the dependence on higher order jets is embedded in the interaction term, specifically, in the potential $1$-form $\bar{A}$. 

\item Third, let us remark that in the theory developed above the extension of the higher order field allows the compatibility of the equation of motion of the point particle with Larmor's radiation law.

\item By imposing the mass renormalization condition \eqref{renormalization of the mass} and the reduction of order condition \eqref{constraint}, the equation \eqref{equationofmotionnoncovariant expression} is not derived directly from an action functional. Such a procedure suggests the phenomenological character of the model of classical charged point particle based upon the equation of motion \eqref{equationofmotionnoncovariant expression}. It also suggests the effective character of the equation, in concordance with previous derivations of the equation by means of formal series of powers of $\epsilon$, as discussed previously \cite{Ricardo2012, Ricardo 2017}.
    \end{itemize}

\section{Energy momentum tensor and balance conditions}\label{Appendix C}
Let us consider the action functional \eqref{action funcional}. Then the Hilbert energy momentum tensor is defined by the relation
\begin{align}
\frac{\delta S}{\delta g_{\mu\nu}}=\int_I \,d\tau\,T^{\mu\nu} (\delta g_{\mu\nu}) ,
\end{align}
where $\tau$ is the proper parameter of $\eta$.
Over a world line $\vartheta :I\to M_4$, the energy momentum tensor can be represented as a distributional density (see for instance the treatment in \cite{Gratus2022}),
\begin{align*}
\mathcal{T}^{\mu\nu}=\,T^{\mu\nu}\,\omega^{-1}\,\delta (x-\vartheta(\tau)),
\end{align*}
where $\omega^{-1}=\,\frac{1}{\sqrt{\det (g)}}$. Then we have the relation
\begin{align*}
\int \mathcal{T}^{\mu\nu}\omega \,d^4 x =\,T^{\mu\nu}(\vartheta(\tau)) .
\end{align*}
Due to the constrain \eqref{constraint}, the equation of motion \eqref{equation of motion} is equivalent to the formal Lorentz force equation,
\begin{align}
m\,(\nabla_{\dot\vartheta}\dot\vartheta)^\mu =\,q\,\bar{F}^{\mu}\,_{\nu}\,\frac{d\vartheta^\nu}{d\tau},
\end{align}
where the covariant derivative is the one associated to the Christoffel symbols of $g$. Such connection is indeed the Berwald connection of $g$, although it is equivalent to the Levi-Civita connection of $\eta$ \cite{Ricardo 2017b}.
The divergence of $\mathcal{T}^{\mu\nu}$ can be expressed as
\begin{align}
\nabla_\nu \mathcal{T}^{\mu\nu}=\,q\,\bar{F}^{\mu}\,_{\nu}\,\frac{d\vartheta^\nu}{d\tau} .
\end{align}
Thus in a region surrounding the world line of a particle, we have the following balance relation,
\begin{align}
\int_U \,\nabla_\nu \mathcal{T}^{\mu\nu} =\,\int_I\,q \bar{F}^{\mu}\,_\nu(\,^2\vartheta(\tau))\,\frac{d\vartheta^\nu}{d\tau} d\tau.
\label{general balance equations}
\end{align}
This is the general form of the energy-momentum balance equations. It has several reads, but the general meaning is that for $\mu=0$, the energy momentum change of the point particle is generated by the external action of the field on the point particle.
At the location of the point particle that serves as a testing particle, the generalized field has two different terms, $\bar{F}=\,F(\vartheta)\,+\Upsilon (\,^2\vartheta)$. Therefore, there are two reads of the balance equation. In the first one, the complete balance equation reads
\begin{align}
\int_U \,\nabla_\nu \mathcal{T}^{\mu\nu}\,\omega\,d^4x =\,q \int_I\,\left({F}^{\mu}\,_\nu\,(\vartheta(\tau)+\Upsilon^{\mu}\,_\nu(\,^2\vartheta)\right)\frac{d\vartheta^\nu}{d\tau}d \tau .
\label{complete balance equation}
\end{align}
It contemplates the changes in stress-energy fluxes due to the radiation of the point particle, which is supported by the term $\Upsilon$ of the generalized field.

Note that since the term $\Upsilon$ serves to compensate the flux of energy due to radiation by the particle. Therefore, in situations when radiation is negligible, the balance equation is
\begin{align}
\int_U \,\nabla_\nu \mathcal{T}^{\nu 0}_{mech}\,\omega\,d^4x =\,\int_I \,q {F}^{0}\,_{\nu}(\vartheta(s))\frac{d\vartheta^\nu}{d\tau}d \tau ,
\label{approximate balance equation}
\end{align}
where $\mathcal{T}^{\nu 0}_{mech}$ is the energy-momentum density of the point particle. By means of Stokes' theorem, the relation \eqref{approximate balance equation} when apply to the action functional \eqref{action funcional 2} leads to the relation
\begin{align}
\mathcal{E}(\tau_2)\,-\,\mathcal{E} (\tau_1)=q\, \Phi(\vartheta(\tau_2))-q\, \Phi(\vartheta(\tau_1)) ,
\label{balance of energy in a spacetime of maximal acceleration}
\end{align}
where the kinetic energy $\mathcal{E}$ is associated to the metric $g$. Note that, according to the definition of kinetic energy \eqref{modifiedE}, there are two contributions to the kinetic energy: the one due to the velocity and the one due to the acceleration. In the theory discussed here, it is this energy the one that appears in the conservation law \eqref{balance of energy in a spacetime of maximal acceleration}.

Compared with the standard balance equation, where the kinetic energy $E$ depends only on the speed ${\bf v}_L$ of the system, the relation is of the form
\begin{align}
{E}(\tau_2)\,-\,{E} (\tau_1)=q\, \Phi(\vartheta(\tau_2))-q\, \Phi(\vartheta(\tau_1)) .
\label{standard balance of energy in a spacetime of maximal acceleration }
\end{align}

\subsection*{Acknowledgements} The author would like to thank Dr. C. Castro Perelman for very interesting correspondence and to the anonymous referees for their suggestions and comments of a previous version of the work.
\small{
}


\begin{thebibliography}{22}

\bibitem{Caianiello} E. R. Caianiello, {\it Is there a Maximal Acceleration}, Lett. Nuovo Cimento {\bf 32}, 65 (1981).

\bibitem{Caianiello 1992} E. R. Caianiello, {\it Quantum and Other Physics as Systems Theory}, La Rivista del Nuovo Cimento, Vol. {\bf 15}, Nr 4 (1992).

\bibitem{Ricardo Nicolini 2018} R. Gallego Torrom\'e and P. Nicolini, {\it Theories with maximal acceleration}, International Journal of Modern Physics A  {\bf 33}, 1830019 (2018).

\bibitem{Ricardo 2019} R. Gallego Torrom\'e, {\it Some consequences of theories with maximal acceleration in laser-plasma acceleration}, Modern Physics Letters A Vol. {\bf 34} (2019) 1950118.

 \bibitem{Ricardo 2024} R. Gallego Torrom\'e, {\it On the effect of the maximal proper acceleration in the inertia}, Proceedings of the Royal Society A {\bf 480}: 20230876 (2024).

\bibitem{Ricardo 2025}	R. Gallego Torrom\'e, {\it On the Unruh effect for hyperbolic observers in spacetimes of maximal proper acceleration}, Proceedings of the Royal Society A {\bf 481} 20240931 (2025).

\bibitem{Ricardo 2025b} R. Gallego Torrom\'e, {\it Spacetimes of maximal proper acceleration and muon mean lifetime}, https://www.researchgate.net/publication/395467866, to appear in International Journal of Geometric Methods in Modern Physics.

\bibitem{Mashhoon1990} B. Mashhoon, {\it The Hypothesis of Locality in relativistic physics}, Phys. Lett. A {\bf 145}, 147–153 (1990).

\bibitem{Einstein1922} A. Einstein, {\it The meaning of relativity}, Princeton University Press (1923).

\bibitem{Pauli 1958} W. Pauli, {\it Theory of Relativity}, Pergamon Press (1958).


\bibitem{Ricardo2007} R. Gallego Torrom\'e, {\it On a covariant version of Caianiello's Model}, Gen. Rel. Grav. {\bf 39}, 1833-1845 (2007).

\bibitem{Ricardo2015} R. Gallego Torrom\'e, {\it An effective theory of metrics with maximal proper acceleration}, Class. Quantum. Grav. {\bf 32}, 2450007 (2015).

    \bibitem{Ricardo 2017} R. Gallego Torrom\'e, {\it A second order differential equation for a point
Charged particle}, Int. J. Geom. Methods Mod. Phys., {\bf 14}, No. 04, 1750049 (2017).

\bibitem{Ricardo2012} R. Gallego Torrom\'e, {\it Geometry of generalized higher order fields and applications to classical linear electrodynamics}, arXiv:1207.3791.

\bibitem{Saunders 1989} D. J. Saunders, {\it The geometry of Jet Bundles}, London Mathematical Society Lecture Note Series 142, Cambridge University Press (1989).

\bibitem{KolarMichorSlovak} I. Kolar, P. W. Michor and J. Slovak, {\it Natural operators in differential geometry}, Springer-Verlag (1993).

\bibitem{Ricardo05b} R. Gallego Torrom\'e,  {\it Quantum systems as results of geometric evolutions}, arXiv:math-ph/0506038.

 \bibitem{Ricardo06} R. Gallego Torrom\'e, {\it A Finslerian version of 't Hooft Deterministic Quantum Models},
J. Math. Phys. {\bf  47}, 072101 (2006).

\bibitem{Ricardo2014} R. Gallego Torrom\'e, {\it Foundations for a theory of emergent quantum mechanics and emergent classical gravity}, arXiv:1402.5070 [math-ph].

\bibitem{Ricardo 2024b} R. Gallego Torrom\'e, {On the emergence of gravity in a framework of emergent quantum mechanics},
International Journal of Geometric Methods in Modern Physics, Vol. {\bf 21}, No. 14, 2450245 (2024).

\bibitem{Ricardo 2021}	R. Gallego Torrom\'e, {\it Average of geometric structures in Finsler spaces with Lorentzian signature}, International Journal of Geometric Methods in Modern Physics, Vol. {\bf 18}, No. 07, 2150107 (2021).

\bibitem{BennTucker} I.M. Benn, R.W. Tucker, {\it An introduction to Spinors and Geometry with applications in Physics}, Bristol ;
Philadelphia : A. Hilger (1987).


\bibitem{Recami} E. Recami, {\it Classical tachyons and possible applications}, Riv. Nuovo Cimento {\bf 9}, 32–33 (1986).


\bibitem{Sauter} F. Sauter, {\it \"Uber das Verhalten eines Elektrons im homogenen elektrischen Feld nach der relativistischen Theorie Diracs},  Zeitschrift für Physik {\bf 69}, 742–764 (1931).

\bibitem{Schwinger} J. Schwinger, {\it On Gauge Invariance and Vacuum Polarization} Phys. Rev. {\bf 82}, 664–679 (1951).

\bibitem{Jackson} J. D. Jackson, {\it Classical Electrodynamics}, Third ed. Wiley (1998).

\bibitem{Mantegna M. 2015} M. Mantegna {\it Revisiting Barry Cox and James Hill’s theory of superluminal motion: a possible solution to the problem of spinless tachyon localization}, Proc. R. Soc. A {\bf 471}: 20140541 (2015).

\bibitem{Born 1938} M. Born, {\it A Suggestion for Unifying Quantum Theory and Relativity}, Proceedings of the Royal Society A {\bf 165}, 291 (1938).

\bibitem{Born 1949} M. Born, {\it Elementary Particles and the Principle of Reciprocity}, Nature {\bf 163} (1949) 207.

\bibitem{Low1993} S. Low, {\it U (3, 1) Transformations with Invariant Symplectic and Orthogonal Metrics} Nuovo
Cimento B {\bf 108} (1993) 841.
\bibitem{Low 2002} S. Low, {\it Representations of the canonical group, the semi-direct product of the unitary and
Weyl-Heisenberg groups, acting as a dynamical group on noncommutative extended phase
space}, J. Phys. A : Math. Gen., {\bf 35} (2002) 5711.

\bibitem{Castro 2005a} C. Castro, {\it Is Dark Matter and Black-Hole Cosmology an Effect of Born Reciprocal
Relativity Theory?}. Canadian Journal of Physics. Published on the web 29 May 2018.

\bibitem{Castro 2005b} C. Castro, {\it Some consequences of Born Reciprocal Relativity in Phase Spaces}, Foundations
of Physics {\bf 35}, no.6 (2005) 971.


\bibitem{Ricardo Paolo Piccione 2015} R. Gallego Torrom\'e and P. Piccione, {\it The Lie group structure of pseudo Finsler isometries}, Houston Journal of Mathematics {\bf 41} (2), 513-521 (2015).

\bibitem{GallegoPiccioneVitorio:2012} R. Gallego Torrom\'e, P. Piccione and H. Vitorio, {\it Fermat principle for timelike curves in Finsler spacetime}, Journal of Mathematical Physics {\bf 53},123511 (2012).

\bibitem{Ricardo 2024c} R. Gallego Torrom\'e, {\it Averaged structures associated with a Finsler structure}, Annals of the "Alexandru Ioan Cuza" University of Iaşi (New Series) Mathematics Vol. LXX, 2024, f. 2.



\bibitem{Castro-Perelman 2025} C. Castro Perelman, {\it Born Reciprocal Relativity Theory Redefines the notion of Mass: How Massless photons appear massive in accelerated frames},  https://www.researchgate.net/publication/389361816 , to appear in IJGMMP.

\bibitem{Castro-Perelman 2025b} C. Castro-Perelman, {\it Born Reciprocal (Non-inertial) Relativity, Phase Space Trajectories and Strings with variable Tension}, https://www.researchgate.net/publication/394998080 .



\bibitem{Rohrlich} F. Rohrlich, {\it The dynamics of a charge sphere of the electron}, Am. J. Phys. {\bf 65}, 1051-1056 (1997).


\bibitem{Bonnor} W. B. Bonnor, {\it A new equation of motion for a radiating charged
 particle}, Proc. R. Soc. Lond. A {\bf 337}, 591-598 (1974).

\bibitem{Nesterenko 1989} V. V. Nesterenko, {\it Singular Lagrangians with higher derivatives}, J. Phys. A: Math. Gen. {\bf 22}, 1673-1687 (1989).

\bibitem{Lovelock Rund} D. Lovelock and H. Rund, {\it Tensors, Differential Forms, and Variational Principles}, Dover (1989).

\bibitem{Gratus2022} J. Gratus, {\it Maxwell-Lorentz without self interactions: conservation of energy and momentum}, J. Phys. A: Math. Theor. {\bf 55} 065202 (2022).

\bibitem{Ricardo 2017b} R. Gallego Torrom\'e, {\it On singular generalized Berwald spacetimes and the equivalence principle},
International Journal of Geometric Methods in Modern Physics, {\bf 14}(06):1750091, 2017.



\end{thebibliography}
\end{document}